\documentclass[aps,prd,twocolumn,groupedaddress,showpacs]{revtex4}
\usepackage{graphicx}
\newcommand{\be}{\begin{eqnarray}}
\newcommand{\ee}{\end{eqnarray}}
\begin{document}
\title{Analytic Results for a Flat Universe Dominated by Dust and
Dark Energy}

\author{A.~Gruppuso, F.~Finelli} 

\thanks{Both the 
authors are also supported by INFN, Sezione di
Bologna, via
Irnerio, 46 -- I-40126 Bologna -- Italy}

\affiliation{
INAF/IASF-BO, Istituto di Astrofisica Spaziale e Fisica Cosmica, 
Sezione di Bologna \\
Istituto Nazionale di Astrofisica, via Gobetti 101, I-40129 Bologna - Italy}

\email{gruppuso@iasfbo.inaf.it, finelli@iasfbo.inaf.it}

\date{\today}

\begin{abstract}
We find the solution for the scale factor in a flat Universe driven by
dust plus a component characterized by a constant parameter of state
which dominates in the asymptotic future.
We also present an analytic formula (in terms of hypergeometric functions)
for the past light cone in such a universe. As applications of this result, we give analytic expressions for the Luminosity Distance and for the Acoustic
Scale, where the latter determines the peaks positions in the
pattern of CMB anisotropies.
\end{abstract}


\maketitle

\raggedbottom
\setcounter{page}{1}
\section{Introduction}
\setcounter{equation}{0}
\label{intro}

The past light cone, defined as the comoving distance $r$
travelled by a light signal from time $t$
to the present time $t_{0}\,(>t)$, in cosmology is given by \cite{KT}
\be
r (t, t_0) = \frac{1}{H_0 \sqrt{|\Omega_K|}} S_K \left( H_0 
\sqrt{|\Omega_K|} 
\int_t^{t_{0}} \frac{d t'}{a (t')} 
\right)
\label{general}
\, ,
\ee
where $a(t)$ is the scale factor, $H_0$ is the
Hubble parameter (defined as $H=d \ln a/dt $) computed at $t_0$ and
$\Omega_K=-K/(H_0^2 a_0^2)$ is the present curvature energy density
($K=1, 0, -1$ for closed, flat, open spatial sections) with $a_0=a(t_0)$.
Depending on $K=1\,,0\,,-1$ respectively, the
function $S_K$:
\begin{displaymath}
S_k (x)= \left\{ \begin{array}{l} \sin(x)  \\ x \\
\sinh(x) \,
\end{array} \right.
\end{displaymath}
contains information on the geometry of space.

For flat spatial sections Eq. (\ref{general}) simplifies to
\be
r (t, t_0) = \int_t^{t_{0}} \, d t'/ a(t')
\, . \label{cd}
\ee
Changing variable of integration from $t$ to the redshift
$z$ (defined as $z=1/a - 1$, with $a(t_{0})=1$),
one obtains
\be
r (z) = \int_0^{z} \, d z' / H(z')
\, . \label{cdz}
\ee
Since $H(z)$ is given by the Friedmann equation (see Sec.\ref{fried}),
Eq.~(\ref{cdz}) allows the computation of $r$ without solving for $a(t)$.
Of course, once $r(z)$ is known, one still has to solve the Friedmann
equation in order to obtain the dependence on time.

The integrations (\ref{cd}),(\ref{cdz}) are the starting points for the computation of
some distances of astrophysical interest as the Proper Distance (or Horizon), the Angular Diameter Distance and the Luminosity Distance \cite{peebles}.
The comoving distance (\ref{cdz}) is also needed for the computation of the Acoustic Scale that is the
characteristic angular scale of the peaks in the Power Spectrum of the Cosmic Microwave Background (henceforth CMB) anisotropies.

An analytic form for $r$ is known for few simple
cases \cite{KT} because is difficult to perform the integration
either in form of
Eq.~(\ref{cd}) or Eq.~(\ref{cdz}). In the case of dust plus curvature,
the time evolution for $a$ and $r$ are known \cite{KT}.

Here we present the analytic solution of the scale factor {\em and} the exact computation of the
past light cone in a flat Universe
driven by Dust (whose energy density $\rho_C(t)\propto a^{-3}$) and Dark Energy ($\rho_X(t)\propto
a^{\alpha}$) with a parameter of state (i.e. pressure over energy)
constant in time. The important case of $\Lambda$CDM is obtained setting
$\alpha =0$. We apply this formula to the following cases:
first, we provide the exact expression of the Luminosity Distance
(and we give an expansion of the past light cone at large $z$) and then
we write down the analytic expression for the Acoustic Scale, relevant for
the peaks analysis of the angular power spectrum of CMB anisotropies.

The paper is organized as follows: in section \ref{fried}
we solve the Friedmann equation introducing a new time coordinate.
In section \ref{plc} we find the analytic formula for the past light cone.
We apply our results to the luminosity distance in section \ref{ldistance} and to the acoustic scale in section \ref{sectionas}.
We conclude in section \ref{conclusion}.

\section{Solution of the Friedmann equation}
\label{fried}
In a flat universe, considering the FRW metric
$ds^2 = -dt^2 + a^2(t) d \vec x^2$,
filled with dust and another component with a constant parameter of state
($w_X = -1 -\alpha/3 $)
we have
\be
\rho_C (t) = \rho_C \, a^{-3}(t)
\mbox{      and        }
\rho_X (t) = \rho_X \, a^{\alpha} (t)
\, ,
\ee
where $\rho_X \,, \rho_C$ do not depend on time and are the energy density
of the two components at present time.
For $\alpha > -2$ the $X-$component drives the universe into asymptotic
acceleration and for
$\alpha > 0$ such component violate the dominant energy condition ($w_X <
-1$). The scale factor $a (t)$ satisfies the Friedmann equation
\be
H^2 \equiv \left( \frac{d a}
{a d t}
\right)^2= {8 \pi G \over {3}} \left[ \rho_X a^{\alpha} + \frac{\rho_C}
{a^3}\right]
\, .
\label{friedeq}
\ee
The total state parameter is
\be
w_{\rm TOT} (a) = - \left(1 + \frac{\alpha}{3} \right)
\frac{\rho_X \, a^{\alpha + 3}}{\rho_X \,
a^{\alpha+3} + \rho_C}
\ee
and interpolates between $0$ and $w_X$.
\noindent
It is not easy to find solutions for a two-component system (see however
the radiation-matter universe and one of its generalizations
\cite{bounce}), but Eq.~(\ref{friedeq}) can be integrated in the
$\alpha =0$ case (i.e. $\Lambda$CDM model) \cite{staro}.
When $\alpha \neq 0$, it is no more possible to obtain
analytically $a=a(t,\alpha)$ but only its inverse $t=t(a,\alpha)$.
This problem can be solved defining a new time coordinate $\tau$:
\be
dt = d\tau \, a^{-\alpha /2}
\, ,
\label{tau}
\ee
such that in this new coordinate, eq.~(\ref{friedeq}) is given by
\be
H^2_{\tau} = {8 \pi G \over {3}} \left[ \rho_X  + \frac{\rho_C}
{a^{3+\alpha}} \right]
\, ,
\label{friedeqintau}
\ee
with $H_{\tau} = a^{\prime}/ a$, where the prime stands for derivative with
respect to the time $\tau$. Note that only for $\alpha = -2$, i.e. when
the $X$-component behaves like a curvature term, $\tau$ becomes the
conformal time $\eta$ (defined as $d \eta = d t/a$).
Eq.~(\ref{friedeqintau}) can be integrated
for $\alpha > -3$ with the result:
\begin{eqnarray}
& & a(\tau) = \left({\Omega_C \over {1-\Omega_C }}\right)^{1/(3+\alpha)} \times \nonumber \\
& & \times \left[ \sinh{\left( {3+\alpha \over{2}} (1-\Omega_C)^{1/2} H_0 \tau
\right)} \right]^{2/(3+\alpha)}
\, ,
\label{solution}
\end{eqnarray}
where $\Omega_C = 8 \pi G \rho_C / 3 H_0^2 = 1 -\Omega_X$ is the present
dust density ratio \cite{bouncing}.
For $\tau \rightarrow 0$, we find the standard dust behaviour $a \sim
\tau^{2/(3+\alpha)} \sim t^{2/3}$.
Notice that when $\alpha \rightarrow 0$, Eq.~(\ref{solution}) recovers the known solution
for $\Lambda$CDM model \cite{staro}.

\section{Past Light Cone}
\label{plc}

In order to perform the integration (\ref{cd}), we make use again of
Eq.~(\ref{tau}) so that is possible to rewrite the comoving distance as
\be
r(a) = \int_a^1 {da \over {a^{1 + \alpha/2} a^{\prime}}}
\, .
\label{cd2}
\ee
By making explicit the form of $H_\tau$ in the integral, Eq.~(\ref{cd2})
reads
\begin{eqnarray}
r(a) &=& \left({1-\Omega_C \over {\Omega_C
}}\right)^{{(\alpha+2)\over{2(\alpha+3)}}}
{1 \over {H_0 (1-\Omega_C)^{1/2}}} \times \nonumber \\
&\times &\int_{a/A}^{1/A} {dx \over {x^{1/2} (1+x^{3+\alpha})^{1/2}}}
\, ,
\label{cd3}
\end{eqnarray}
where $1 / A = {(-1 + \Omega_C^{-1})}^{1/(3+\alpha)}$.
The integration left in Eq.~(\ref{cd3}) can be performed giving
(see $15.3.1$ of \cite{AS})
\begin{eqnarray}
r(z) &=&
{1\over {H_0}}
{2 \over {\Omega_C^{1/2}}}
\left[ f \left(\alpha,{1 \over {\Omega_C}}-1\right) 
\right. 
\nonumber 
\\  & & \left. - \left({1\over{1+z}}\right)^{1/2} f\left(\alpha,{1-\Omega_C \over {\Omega_C (1+z)^{3+\alpha}}}  
\right) \right] \, ,
\label{mainformula}
\end{eqnarray}
where the definition of the redshift $z$ has been used and where $f$ is the following hypergeometric function \cite{AS}
\begin{eqnarray}
f(\alpha,y) &=& \, _2 F_1\left[{1\over{2(\alpha+3)}},{1\over 2}, 1+
{1\over{2(\alpha+3)}},-y\right] \nonumber \\
&=& \frac{(-y)^{-1\over(6+2\alpha)}}{(6 + 2
\alpha)} B_{-y} \left(\frac{1}{6 + 2 \alpha},
\frac{1}{2} \right)
\, ,
\end{eqnarray}
and in the last equality we have also rewritten the
hypergeometric function in terms of incomplete Beta function
$B_x (\gamma,\beta)$ \cite{AS}.

It is interesting to expand the past light cone for large $z$, obtaining
\begin{eqnarray}
& & \!\!\!\!\!\!\!\!\!\!\!\!\!r (z) = { 2 \over {\Omega_C^{1/2}} H_0} \, \left\{
_2 F_1\left[{1\over{2(\alpha+3)}},{1\over 2}, 1+
{1\over{2(\alpha+3)}}, \right. \right. \nonumber \\
& & \!\!\!\!\!\!\!\!\!\!\!\!\! \left. \left. - {1-\Omega_C \over {\Omega_C}}
\right] - \frac{1}{(1+z)^{1/2}} + {\cal O} \left( \frac{1}{(1+z)^{\alpha + 7/2}}
\right)
\right\}
\, ,
\label{largezexpansion}
\end{eqnarray}
where the leading term (that is constant with respect to the redshift $z$)
is the conformal age of the universe already given in \cite{DSW}.
Note that the $\Lambda$CDM cosmic age in cosmic time is given in 
\cite{KT}.

\section{Luminosity Distance}
\label{ldistance}

Eq.~(\ref{mainformula}) is the starting point for the computation of
distances of astrophysical interest
such as the Luminosity and the Angular Distances.
The Luminosity Distance, $d_L$, and the Angular
Distance, $d_A$, in a flat universe are given by
\be
d_L(z) = (1+z) \, r(z)
\label{luminositydistance}
\, ,
\ee
\be
d_A(z) = (1+z)^{-1} \, r(z)
\label{angulardistance}
\, .
\ee

Let us focus on the Luminosity Distance. From
Eq.~(\ref{luminositydistance}), $d_L$ can be expanded for small $z$
by using its form in terms of incomplete Beta function
\be
H_0 d_L(z) = z + d_2 z^2 + d_3 z^3 + d_4 z^4 + O(z^5)
\, ,
\label{expansion}
\ee
with
\be
d_2={(4+\alpha)-(3+\alpha)\Omega_C  \over 4}
\, ,
\ee
\begin{eqnarray}
d_3 &=& {1\over 24}\left[ \alpha (4+\alpha) - 2 (3+\alpha) (5+2\alpha) \Omega_C + \right. \nonumber \\
& & \left. + 3 (3+\alpha )^2\Omega_C^2 \right]
\, ,
\end{eqnarray}
\begin{eqnarray}
d_4 &=& 
{1\over 192} \left[ (-2+\alpha)\alpha (4+\alpha) 
\right. 
\nonumber 
\\ 
& & \left. -(3+\alpha)(40+\alpha (50+13\alpha )) \Omega_C
\right. 
\nonumber 
\\
& & \left. + 3 (3+\alpha)^2 (20+9\alpha)\Omega_C^2 -15 (3+\alpha)^3 \Omega_C^3 \right] \, .
\end{eqnarray}
Once $\alpha$ is set to zero, $d_2$ and $d_3$ recover the known coefficients
in flat $\Lambda$CDM model \cite{visser} with the deceleration parameter
$q_0 = \Omega_C/2 - \Omega_X = 3\Omega_C/2 -1$ and with the jerk parameter
given by $j_0 = \Omega_C + \Omega_X =1$.


\begin{widetext}

\begin{figure}


\begin{tabular}{cc}

\includegraphics[width=8.4cm]{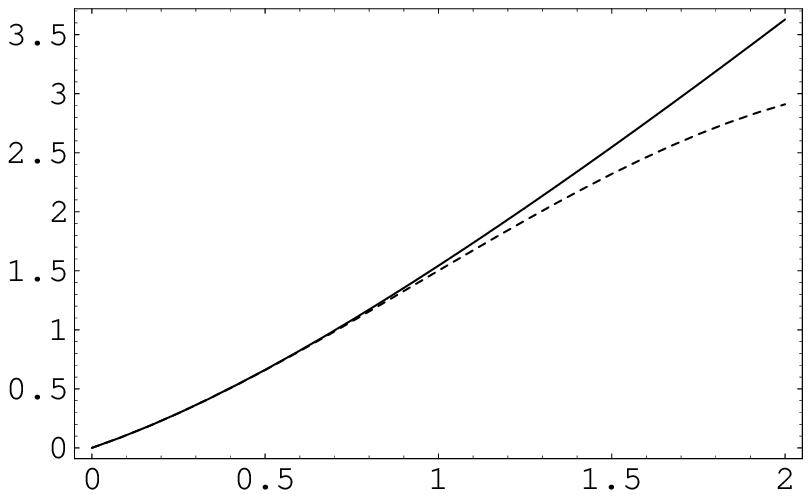}

&

\includegraphics[width=8.4cm]{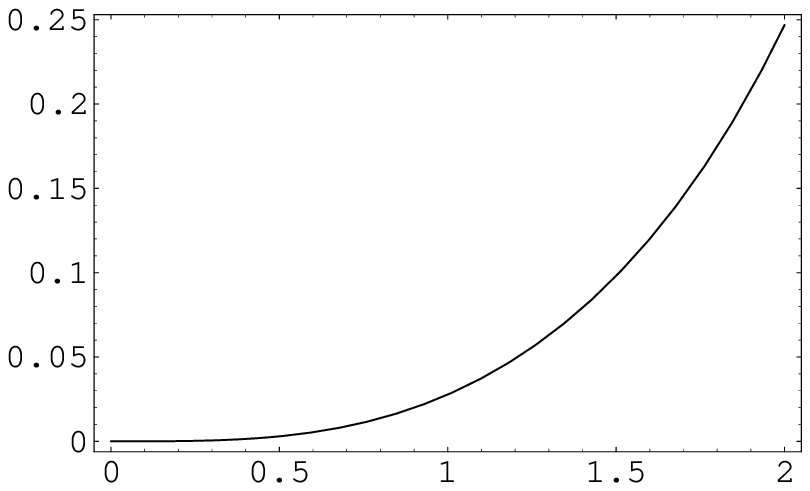}

\end{tabular}

\caption{Left Panel: Exact luminosity distance $H_0 \, d_L$ (solid line) and
approximated luminosity distance $H_0d_{app}$ (up to third order, dotted
line), both vs redshift $z$. Right Panel:
$p$ vs $z$ where $p$ is implicitly defined by $d_L=(1+p) d_{app}$
where $d_L$ is the exact luminosity distance and $d_{app}$ is the approximated
(up to third order) luminosity distance.
In both panels $\Omega_C=0.3$ and $\alpha=0$.}
\label{fig:1}

\end{figure}

\end{widetext}

In Fig.~{\ref{fig:1}} we show the comparison between the exact
(see Eq.~(\ref{luminositydistance})) and the approximated
(up to the third order, see Eq.~(\ref{expansion}))
Luminosity Distance as a function of z for $\Omega_C=0.3$ and $\alpha=0$.
For istance at $z=1.5$, an error of $9.8 \%$ is made, if one considers
the approximated instead of the exact luminosity distance.

\begin{figure}

\begin{center}

\begin{tabular}{c}
\includegraphics[width=84mm]{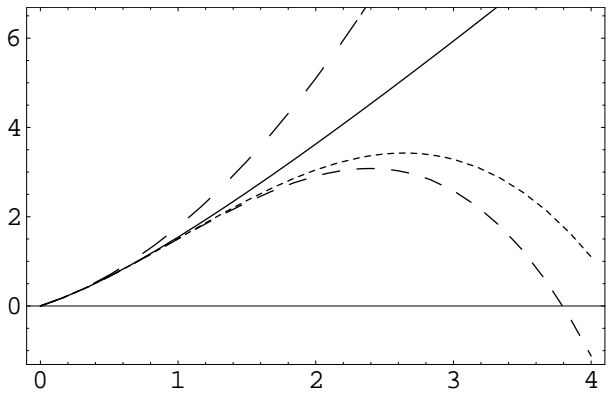}
\end{tabular}

\caption{Exact luminosity distance (solid line) and three approximated
luminosity distance
for $\Omega_C=0.3$ and $\alpha=0$ vs redshift.
Long, short dashed and dotted line are second, third and fourth order in
$z$ expansion, respectively. Future cosmological interpretations of 
high-redshift data \cite{hd} need the accuracy of the exact solution.}
\label{fig:2}

\end{center}

\end{figure}

In Fig.~{\ref{fig:2}} we compare the exact luminosity distance
with the first three approximated luminosity distances
in a flat $\Lambda$CDM model.
We note that the approximation of order $z^2$ always overestimate the
exact luminosity distance while the approximations of order $z^3$ and
$z^4$ always underestimate the exact luminosity distance. From this
analysis it is clear that already for $z \sim 2$ the use of a third order
expansion may flaw the comparison of observations with theoretical
predictions.

\begin{figure*}

\begin{center}

\begin{tabular}{cc}
\includegraphics[width=8.4cm]{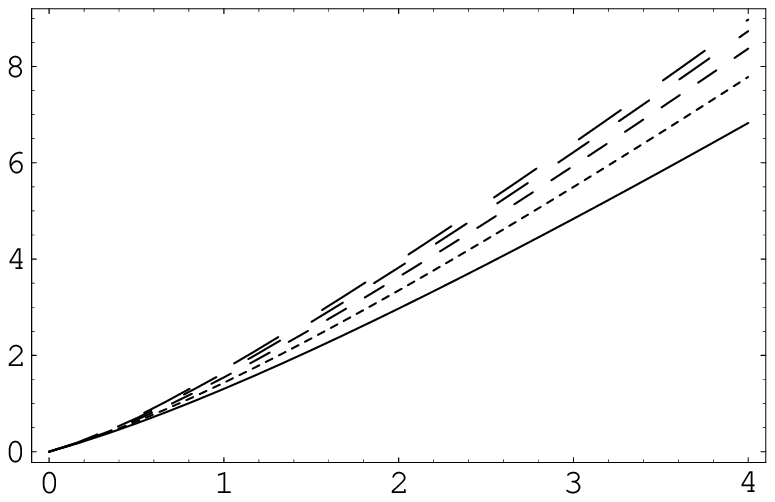} &
\includegraphics[width=8.4cm]{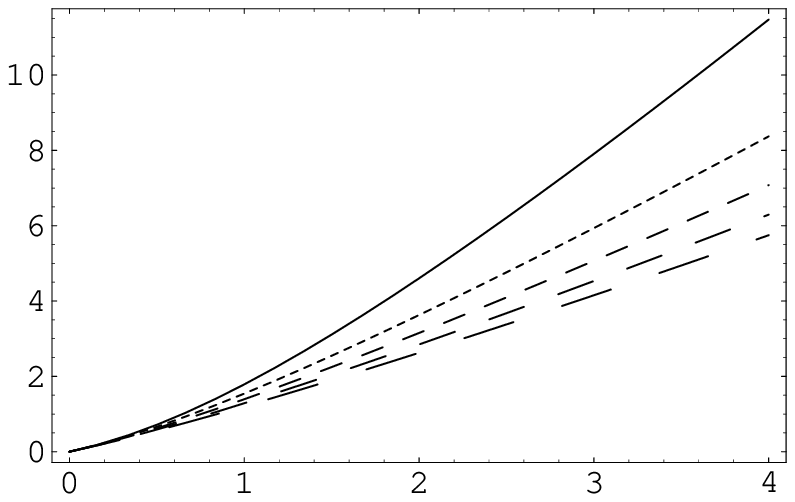}
\end{tabular}

\caption{Left Panel: Exact luminosity distance vs redshift for $\Omega_C=0.3$ and $\alpha = -2$ (solid line), $\alpha = -1$ (dotted line),
$\alpha = 0$ (short dashed), $\alpha = 1$ (dashed line) and $\alpha = 2$ (long dashed line).
Right Panel: Exact luminosity distance vs redshift for $\alpha =0$ and for $\Omega_X/\Omega_C = 9$
(solid line), $\Omega_X/\Omega_C = 7/3$
(dotted line), $\Omega_X/\Omega_C = 1$ (short dashed line), $\Omega_X/\Omega_C = 3/7$ (dashed line),
$\Omega_X/\Omega_C = 1/9 $ (long dashed line).}
\label{fig:3}

\end{center}
\end{figure*}

In the left panel of Fig.~{\ref{fig:3}} we set $\Omega_C = 0.3 $ (and hence $\Omega_X=0.7$)
and plot Eq.~(\ref{luminositydistance}) for different values of $\alpha$.
We notice that the dependence on $\alpha$ is quite weak.
In the right panel of Fig.~{\ref{fig:3}} we set $\alpha =0$ (i.e. $\Lambda$CDM model)
and plot Eq.~(\ref{luminositydistance})
for different values of the ratio $\Omega_X/\Omega_C$.
Of course larger is this ratio and larger is the luminosity distance for each fixed redshift.
We notice that the dependence on this ratio is stronger than the $\alpha$ dependence.

\section{Acoustic Scale}
\label{sectionas}

The characteristic angular scale of the peaks of the angular power spectrum in CMB anisotropies is defined \cite{WMAPpage} as
\be
\theta_A = {\pi \over {\ell_A}} = {r_{s}(z_{dec}) \over r(z_{dec})}
\label{as}
\, ,
\ee
where $r_{s}(z_{dec})$ is the comoving size of the sound horizon at decoupling, $r(z_{dec})$ is the comoving distance at decoupling given in Eq.~(\ref{mainformula}) and $\ell _A$ is the multipole associated to the angular scale
$\theta_A$, also called acoustic scale.
In fact, the expression (\ref{mainformula}) misses the contribution coming 
from the {\em radiation} ($\Omega_{rad}$).
A numerical study shows that we get an overestimate for $\ell_A$ that depends on $\alpha$
and in the worst case it is of the order of $0.5 \%$.
The comoving size of the sound horizon is given by
\be
r_{s}(z_{dec}) = \int_0^{1/(1+z_{dec})} \, {d a \over {a^2 H(a)}} \, c_s(a)
\, ,
\label{rs}
\ee
where $c_s$ is the sound speed given by \cite{huandsugiyama}
\be
c_s(a)= 1/ \sqrt{3+{9 \over 4} {\Omega_b \over \Omega_{\gamma}}a}
\, ,
\ee
where $\Omega_b$ and $\Omega_{\gamma}$ are respectively, the present density ratio for baryons and for photons.
The integral (\ref{rs}) can be analytically performed once $\Omega_X $, that enters in $H(a)$, is set to zero.
The result is \cite{huandsugiyama}
\begin{eqnarray}
 r_{s}(a_{dec}) &=& {4 \over 3 H_0} \sqrt{\Omega_{rad} \over {\Omega_C \Omega_b}} \times \nonumber \\ 
& & \times \ln \left[ {\sqrt{1+ R_{dec}} + \sqrt{R_{dec}+R_{eq}}\over{1+\sqrt{R_{eq}}}}
\right] \, ,
\label{rscomputed}
\end{eqnarray}
with $R(a)= 3 (\Omega_b/ (4 \Omega_{\gamma})) a$ and
where the label ${dec}$ stands for ``computed at decoupling'' while ${eq}$ stands for ``computed at equivalence''(between radiation and matter).
A numerical study shows that in the worst case this approximation (i.e. $\Omega_X =0$ in $r_{s}$) gives an error of the order of $10^{-5}\,\%$, then completely negligible.

We are now ready to make a prediction for the acoustic scale $\ell_A$ defined in Eq.~(\ref{as}).
We consider $\alpha = 0$ (i.e. $\Lambda$CDM) and set $a_{dec}=1/1089$, $\Omega_C =0.27 $, $\Omega_{b}=0.046$ and $\Omega_X = 0.73$.
In order to compute $z_{eq}$ we made use of the following relation \cite{WMAPpage}
\be
1+z_{eq} = {5464 \, (\omega_m/0.135) \over {(T_{CMB}/2.725)^4 (1+\rho_{\nu}/\rho_{\gamma})}}
\, ,
\label{zeq}
\ee
where $\omega_m = \omega_b + \omega_{cdm} = \Omega_C h^2$ (with
$h=H_0/(100 \, {\rm km} \, s^{-1} \, {\rm Mpc}^{-1}$),
$\rho_{\nu}/\rho_{\gamma}$
is the ratio of neutrinos
to photons densities and $T_{CMB}$ is the CMB temperature in $^0 K$.
Setting $T_{CMB}=2.725$, $h=0.72$, $\rho_{\nu}/\rho_{\gamma}=0.6851$
\cite{WMAPpage} we get $z_{eq}=3362$.
Once $z_{eq}$ is known, it is possible to compute $\Omega_{\gamma}$ through
\be
\Omega_{\gamma} = {\Omega_C \over {(1+\rho_{\nu}/\rho_{\gamma})(z_{eq}+1)}}
\, ,
\label{omegagamma}
\ee
obtaining $\Omega_{\gamma} = 4.76 \, \, 10^{-5}$.
Thus we have all we need to compute the acoustic scale defined in Eq.~(\ref{as}).
We obtain
\be
\ell_A = 301.76
\label{valueforas}
\, ,
\ee
which is in agreement with the Wilkinson Microwave Anisotropy Probe (WMAP) $1 \sigma$
result \cite{WMAPpage}: $ \ell^{WMAP}_A = 300 \pm 3 $.

So our analytic calculation of the acoustic scale agrees
with the value reported by the WMAP team. This is a further application of 
our results to a
different set of observations, i.e. CMB anisotropies, which require a
correct evaluation of the past light cone at high redshift, where a
perturbative expansion does not work and only numerical methods were used
so far.

\begin{figure}

\begin{center}

\begin{tabular}{c}
\includegraphics[width=84mm]{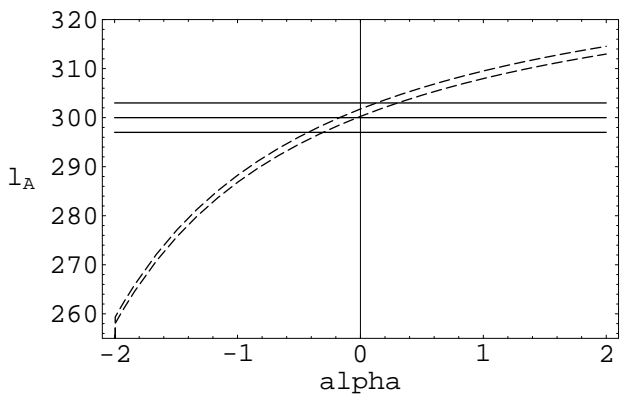}
\end{tabular}

\caption{Acoustic Scale vs $\alpha$.
Our analytic evaluation of the acoustic scale lies between the dashed
lines.
The higher dashed curve is computed with Eq.~(\ref{as}) while the lower is the $99.5 \%$ of the previous one.
In this way we give an estimate of the order of magnitude of the error it is made, neglecting
the contribution of the radiation in Eq.~(\ref{mainformula}).
Horizontal lines represent the $1 \sigma$ contour to the value given by
\cite{WMAPpage}.}
\label{fig:4}

\end{center}

\end{figure}

Therefore, by using Eq.~(\ref{as}) and Eq.~(\ref{largezexpansion}), the acoustic scale is given by: 
\begin{widetext}
\be
\ell_A \simeq {3 \pi \over {2}} \sqrt{\Omega_b \over \Omega_{\gamma}}
\,  {_2 F_1\left[- 1/ (6\omega_X), 1/2, 1- 1/(6\omega_X),
- (1-\Omega_C)/ \Omega_C \right] -1/{(1+z_{dec})^{1/2}}\over 
{\ln \left[ \sqrt{1+ R_{dec}} + 
\sqrt{R_{dec}+R_{eq}} \right] - \ln
 \left[ {1+\sqrt{R_{eq}}} \right]} }
\label{lalargez} \, ,
\ee
\end{widetext}
where the dependence on the cosmological parameters and in particular on 
$\omega_X$ is explicit. Our Eq.~(\ref{lalargez}) is completely analytic 
compared to $\Delta \ell$ given in \cite{doran}, in which 
$r(z_{dec})$ contains a function $F$ that is computed numerically 
\footnote{The function $F$ becomes the hypergeometric function given in 
eq.~(\ref{largezexpansion}) when the contribution coming from radiation is neglected.} and $r_{s}(z_{dec})$ contains an averaged speed of sound that is fixed to a particular numerical value.
In Ref.~\cite{doran} the dark energy parameter of state is a function of 
time, but is taken as a time-average for the final expression of $\Delta 
\ell$ in the models considered in \cite{doran}.
In Fig.~{\ref{fig:4}} we show the dependence
of the acoustic scale on $\alpha$, keeping the other parameters fixed to
the values reported above in the text.

\section{Conclusion}
\label{conclusion}
We have found the analytic solution for the scale factor {\em and} past
light cone for a flat universe driven by dust and another $X$ component
with a parameter of state constant in time which dominates in the future.
If $w_X < -1/3$, $X$ is the simplest possible parametrization for Dark
Energy.
While Eq. (\ref{solution}) generalizes the previously known solution for
$\Lambda$CDM \cite{staro} to $\alpha \ne 0$, the result
(\ref{mainformula}) extends \cite{DSW} for finite $z$, which is important 
for accurate quantitative predictions.
 

Eq.~(\ref{mainformula}) has been applied to the following cases: the 
computation of
the luminosity distance, relevant for supernovae, and the acoustic
scale, which determines the position of the peaks in the pattern of CMB
anisotropies. Both of these results are original.
For the second case, our analytic prediction is in agreement with
the value reported by the WMAP team.
Note that this latter application requires a
correct evaluation of the past light cone at high redshift, where a
perturbative expansion does not work and only numerical methods were
used so far.

A parameter of state constant in time cover a wide range of cosmological models which include dust.
From a fundamental physics point of
view, $X$ corresponds to a cosmic string network or a
curvature term when $\alpha=-2$ or to a domain walls background when $\alpha=-1$.
A quintessence model with an exponential potential which does not have
tracking behaviour leads to a component with $ -2 < \alpha < 0$ (for
$\alpha > 0$ one has to reverse sign of the kinetic term of a canonical
scalar field, i.e. consider a phantom with an exponential potential).

\end{document}